\documentclass[%
reprint,
amsmath,amssymb,
aps,pre
]{revtex4-2} 
\usepackage{graphicx}% Include figure files
\usepackage{dcolumn}% Align table columns on decimal point
\usepackage{bm}% bold math
\usepackage[mathlines]{lineno}% Enable numbering of text and display math
\usepackage{amsmath}
\usepackage{multirow}
\begin{document}
	
	\title[title]{On the evolution of the mass density profile of dense molecular clouds}
	
	\author{S. Donkov}
	\email{sddonkov@astro.bas.bg}
	\affiliation{Institute of Astronomy and NAO, Bulgarian Academy of Sciences, 72 Tzarigradsko Chausee Blvd., 1784 Sofia, Bulgaria}
	\author{I. Zh. Stefanov}
	\email{izhivkov@tu-sofia.bg}
	\affiliation{Department of Applied Physics, Faculty of Applied Mathematics, Technical University-Sofia, 8 Kliment Ohridski Blvd., Sofia 1000, Bulgaria}
	\author{T. V. Veltchev}
	\affiliation{Faculty of Physics, University of Sofia, 5 James Bourchier Blvd., 1164 Sofia, Bulgaria}
	\affiliation{Universit\"at Heidelberg, Zentrum f\"ur Astronomie, Institut f\"ur Theoretische Astrophysik, Albert-Ueberle-Str. 2, 69120 Heidelberg, Germany}
	\author{R. S. Klessen}
	\affiliation{Universit\"at Heidelberg, Zentrum f\"ur Astronomie, Institut f\"ur Theoretische Astrophysik, Albert-Ueberle-Str. 2, 69120 Heidelberg, Germany}
	\affiliation{Universit\"{a}t Heidelberg, Interdisziplin\"{a}res Zentrum f\"{u}r Wissenschaftliches Rechnen, Im Neuenheimer Feld 225, 69120 Heidelberg, Germany}
	\affiliation{Harvard-Smithsonian Center for Astrophysics, 60 Garden Street, Cambridge, MA 02138, U.S.A.}
	\affiliation{Elizabeth S. and Richard M. Cashin Fellow at the Radcliffe Institute for Advanced Studies at Harvard University, 10 Garden Street, Cambridge, MA 02138, U.S.A.}
	
	% Ivan 0000-0001-8541-252X
	% Ralf 0000-0002-0560-3172
	
	\date{\today}% It is always \today, today,
	
	\begin{abstract}
		We set ourselves the goal of obtaining the equations which govern the evolution of the mass density profile of a dense irrotational molecular cloud. We base our study on the notion of "ensemble of molecular clouds", introduced in our previous work. We model the studied clouds making use of the "ensemble abstract representative member" - a spherically symmetric and isotropic cloud. This cloud is isothermal and radially accreting matter from its surroundings. Starting from the equations of hydrodynamics applied to a self-gravitating isothermal spherical gas cloud, we obtain a system of two first-order non-linear partial differential equations, which govern the evolution of two unknown fields: the exponent of density profile and the accretion velocity. Under the assumption of steady state flow, we get approximate solutions, using the method of leading order terms. Far from the cloud's centre we obtain density profile $\varrho=\ell^{-2}$, and the accretion velocity is constant, while near to the centre we have $\varrho=\ell^{-3/2}$ and $v_{\rm a}\propto\ell^{-1/2}$. Through our dynamical equations, the obtained solutions coincide completely with the solutions found by using the equation of energy conservation of a fluid element (in our previous work). Also, combining the equations of energy balance for a fluid element, we arrive at the conclusion that the cloud's layers, far from the centre, are in a stable dynamical state if the accretion velocity flow is sub- or trans-sonic, otherwise they are marginally stable (moderately supersonic flow) or unstable (supersonic flow). Both solutions are consistent only if the accretion flow is subsonic, and hence the outer layers are stable. Finally, under the assumption that both the accretion velocity and density scale with $\ell$ and their power-law exponents are position-independent, we show that the density scaling exponent far away from the centre is $p=2$ and this value is an attractor. Hence this value should be observable in dense molecular clouds.		
	\end{abstract}

%\physh{Your PhySH keywords here}	
%	\keywords{ISM}
	%Use showkeys class option if keyword
	%display desired
	\maketitle
		
	\section{Introduction}
	\label{sec:Introduction}
	
Molecular clouds (MCs) are cold and dense gas entities, which contain a considerable part of the gaseous mass of a galaxy. In our Galaxy they are usually located in the arms and in the Central Molecular Zone. They are sites of star formation \cite{Elme_Scalo_04,HF_12,KG_16}, so their study is of great importance for understanding the structure and evolution of Galaxy.

MCs consist mostly of molecular hydrogen and,  in our Galaxy, one percent of dust particles, where the most metals are locked \cite{HF_12,KG_16}. They are cold structures with typical temperatures $\sim 10-30 {\rm K}$ and nearly isothermal equation of state \cite{Ferriere_01}. Their physics is complex due to (supersonic compressible) turbulence with Mach numbers ranging from less than 1 to more than 10. Besides the (supersonic) turbulence the main "forces" that govern physical processes in MCs are gravity (self-gravity and tidal forces from surrounding matter), gas-pressure gradients, thermodynamics, magnetic fields, and cosmic rays \cite{Wang_ea_2010,Federrath_15,Mathew_Federrath_21}. This complex physical picture can be explored observationally, by numerical simulations, and by theoretical models. The presented contribution is an attempt to develop a model of the densest parts of MCs (or according to our terminology - "dense molecular clouds"), where the very process of star formation takes place. Our model is based on the one-point statistics of the mass density field, the so called probability density (or distribution) function (PDF). This distribution is an imprint of the structure and evolutionary state of MCs, although a part of information (their morphology and related to it dynamics) for real structure and physics is lost. Nevertheless, the fractal nature of MCs (manifested through the mass scaling law of their substructures) \cite{HF_12,LAL_2010} being their main structural characteristic is encoded in the PDF \cite{Elme_Scalo_04,HF_12,KG_16}. Since the PDF reflects the evolutionary stage of a MC, its form is different during the earlier and the later periods of MC's life. At earlier stages supersonic turbulence dominates cloud's physics and the PDF is nearly log-normal \cite{VS_94,Pad_ea_06,Federrath_ea_10,Molina_ea_12}: $P(s) \propto \exp(-(s+\sigma^2/2)^2/2\sigma^2)$, where $s=\ln(\rho/<\rho>)$ is the log-density and $\sigma$ is the dispersion. At later stages from the high density wing of log-normal distribution arises a power-law tail (hereafter PL-tail), which reads: $P(s) \propto \exp(qs)$, where $q$ is a negative number, typically, in the range: $-3 < q < -1$ \cite{Klessen_2000,KNW_11,Girichidis_ea_14,Veltchev_ea_19,Khullar_ea_21,Schneider_ea_2022}. The latter distribution, in log-log coordinates, is a straight line with a slope $q<0$. This PL-tail is a probabilistic imprint of the densest parts of a MC, where cores (which we label as "dense molecular clouds") are located. It corresponds to a physics dominated by interplay between self-gravity and gas-pressure gradients \cite{Girichidis_ea_14,Li_18,DS_19,VS_ea_19}. To complete the picture we must say that at the latest stages of MC's evolution a second PL-tail with slope $\approx -1$ emerges \cite{KNW_11,Veltchev_ea_19,Schneider_ea_2022,DSVK_24}, corresponding to the densest parts of protostellar cores, namely the vicinity of the centre, where individual protostars or the protocluster is located. Due to the nearly spherical symmetry and isotropy of cores the density distribution in them can be described by simple radial density profile (or power-law function): $\rho \propto l^{-p}$, where $l$ is the radius of a given shell and the exponent $p>0$ is connected to the PL-tail slope through the following simple relation: $p=-3/q$ \cite{FK_13,Girichidis_ea_14,DVK_17}. The above mentioned density profile(s) (or PL-tail PDF(s)) are the main theoretical characteristic of the cores, and their slope(s) reflect(s) the physics and evolutionary stage of them. So studying the physical nature and evolution of $p$ is a task of great importance for understanding the evolution of cores and what physics determines it.

In 1969 Larson \cite{Larson_69} solved numerically the equations governing the collapse of an isolated gas sphere with a mass slightly larger than its Jeans mass. He obtained that far from the centre the mass density profile arrives at a stable slope of $-2$. The same result, also for an isolated gas ball, is obtained numerically by Penston \cite{Penston_69a}. This so called Larson-Penston solution describes the isothermal collapse of homogeneous ideal gas spheres initially at rest. The main forces in their calculations are self-gravity and isothermal gas pressure. Later, in 1977, Shu \cite{Shu_77} studied analytically the unstable hydrostatic equilibrium (gas pressure against self-gravity) and the subsequent collapse of an isothermal sphere with a super-Jeans mass (Shu’s solution describes the so-called "inside-out collapse"). He obtained that the profile of the layers (which are in equilibrium) far from the singularity (in the centre) has a slope of $-2$, while the matter that is falling into the centre is characterized by profile with a slope of $-3/2$ (free-falling). In the same year, Hunter explored this problem extensively \cite{Hunter_77}. Whitworth \& Summers \cite{Whitworth_Summers_85} showed, in 1985, that the Larson-Penston's and Shu's solutions reside at the extreme ends of a whole family of similarity solutions. The discussed works belong to classical dynamical theory for star formation, focused on the interplay between self-gravity on the one side and pressure gradients on the other, considering collapse of an isolated gas cloud. They do not regard other factors, as for example, the accretion from the cloud's surroundings or the conservation of angular momentum and magnetic flux during collapse. These works and many others are summarized by Mac Low \& Klessen in their review \cite{Mac Low_Klessen_2004}. Later, in 2011, Kritsuk, Norman \& Wagner \cite{KNW_11} studied numerically the collapse of isothermal turbulent self-gravitating cloud of size $5~$pc, and at about a half free-fall time obtained PDF, which consists of log-normal part at low densities and two PL-tails with slopes of $-1.7$ and $-1$, accordingly, at high densities. By assuming only self-gravity (i.e. a free-fall) and by using approximate theoretical calculations for the collapse of a sphere, Girichidis et al. \cite{Girichidis_ea_14}, in 2014, obtained that the PDF of mass density converges to PL-tail with a slope of $\approx -1.5$.  More recently, in 2015,  Naranjo-Romero, Vazquez-Semadeni \& Loughnane \cite{N-R_ea_15}, using numerical simulations, investigated a collapsing core embedded in a larger medium (called "the cloud") and accreting material from the latter. They also obtained a density profile of $-2$ in the outer layers of the core during its collapse within the cloud. In 2018 Li \cite{Li_18} obtained a density profile with a slope of $-2$ when gravity, accretion and turbulence interact, during a steady-state gas flows in a spherically symmetric dense cloud. He claims that this slope is universal for scale-free gravitational collapse and that an isothermal state is not a necessary condition. Then, in 2018 and 2019, Donkov \& Stefanov \cite{DS_18,DS_19} investigated the density profile of a radially accreting isothermal gas ball with a very small and dense core located in the centre. Under the assumption of steady-state flow, by using the equation of energy conservation of a fluid element, they obtained two solutions for the density profile. Far from the centre the solution has a slope of $-2$, while near to the centre the density profile has a slope of $-3/2$ (i.e. a free-falling). The result far from the centre was confirmed later, in 2022 by Donkov et al. \cite{DSVK_22}, by using the virial theorem (again under the assumption of steady-state flow). In 2020, Joupart \& Chabrier \cite{Jaupart_Chabrier_2020} explored analytically collapse and dynamics of gravoturbulent molecular cloud, in particular the evolution of density PDF. They reported the appearance of two power law tails in the PDF, with two characteristic exponents, corresponding to two different stages in the balance between turbulence and gravity.  Finally, in 2021 Gomez, Vazquez-Semadeni, \& Palau \cite{Gomez_ea_2021} studied the evolution of an accreting prestellar core and arrived at the result that slope of $-2$ is a stationary solution and it is an attractor for the time dependent solutions. In contrast to us, they assumed that the accreting fluid is in free fall, i.e. gas pressure does not matter. Hence their equation of evolution of the density power exponent is different to that obtained by us.

There are confirmations also from observations. In systems such as molecular clumps forming star clusters, the slope of radial density profile is very close to $-2$ \cite{Mueller_ea_02,Evans_03,Wyr_ea_12,Wyr_ea_16,Palau_ea_14,Csengeri_ea_17,Zhang-Li_17}.

In this contribution we derive the equations, which govern the evolution of the mass density profile $p$ (and hence the evolution of the PL-tail slope $q$), during the physical evolution of the cloud. The obtained first-order non-linear partial differential dynamical equations have approximate solutions, under the assumption of a steady state flow, in two regimes: far from the cloud's centre and near to it. The density and velocity profiles we get, in both cases, coincide with those reported in \cite{DS_19} and obtained by use of the energy conservation law of the fluid element. Comparing the conditions for the energy coefficients that we arrive at in this work and in \cite{DS_19}, one can get conclusion for the stability of the cloud's layers far from the centre, in the different accretion velocity  regimes (sub-, trans-, or supersonic). Also, we present a proof that the steady state value $p=2$ ($q=-1.5$), for the layers far from the cloud's centre, is an attractor, which in turn means that the steady state solution must be the most probable observable stage of the dense clouds.

The paper has the following structure. In section \ref{sec-The model} we present the model of the studied object, which is first introduced in 2018, by Donkov \& Stefanov \cite{DS_18}. Section \ref{sec-Evolution of the density profile} is the main contribution, where in \ref{subsec-equations of the medium} we introduce the equations of the medium, according to the model; in \ref{subsec-equations governing the evolution of the density profile} we obtain the general form of equations governing the evolution of mass density and accretion velocity profiles; in \ref{subsec-analytical solution in the case of steady-state flow} we obtain analytical solutions for the density and accretion velocity profiles in the case of steady state flow; finally, in \ref{subsec-the steady-state solution is an attractor} we present a proof that the steady state exponent $p=2$ is an attractor. Then, in section \ref{sec-discussion}, we refer to previous works and compare our results to some related papers. The paper ends with section \ref{sec-summary}, where we summarize our work.

\section{The model}
\label{sec-The model}

\begin{figure}
	\includegraphics[width=0.4\textwidth]{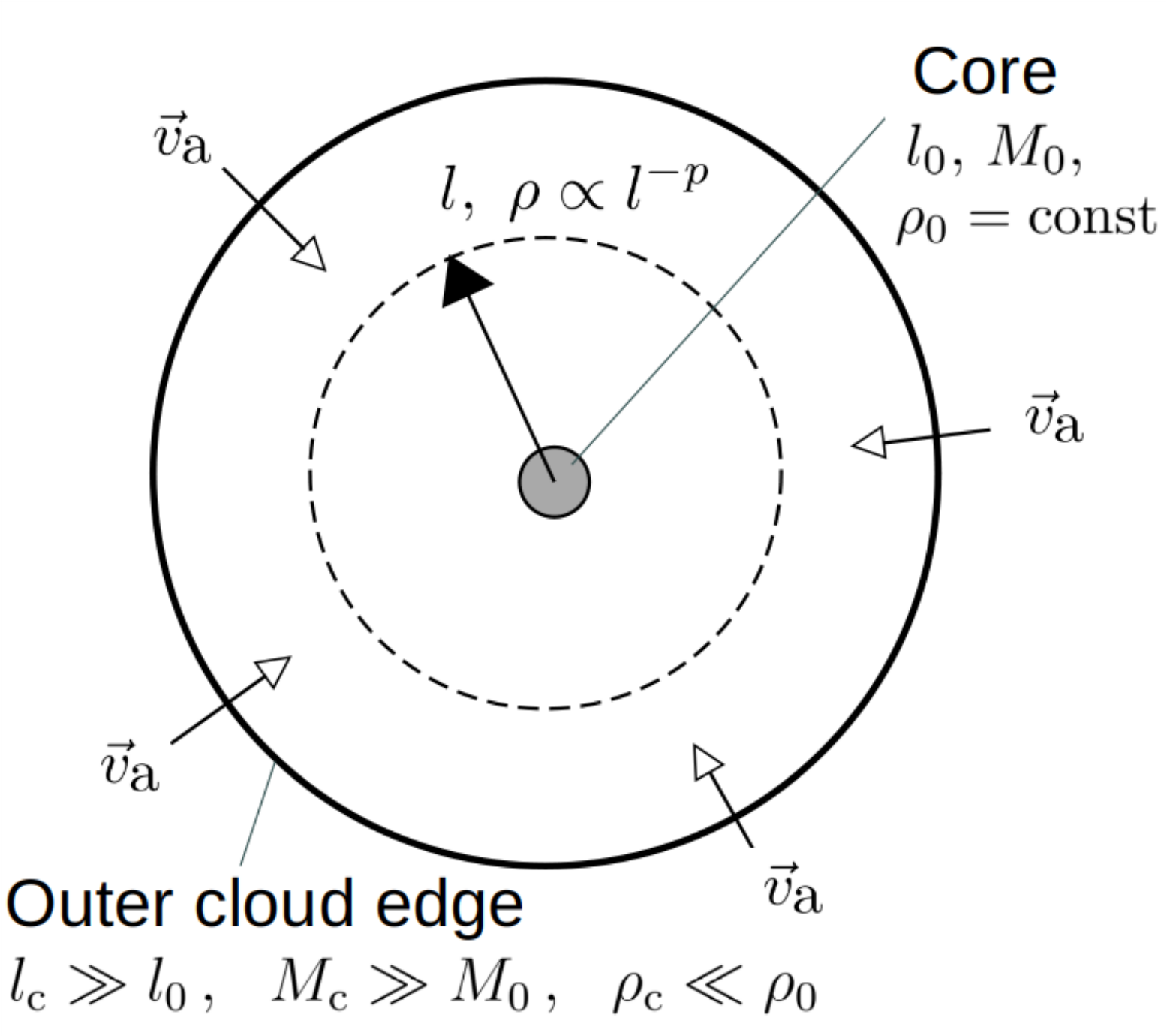}
	\caption{\label{fig:Fig_1_DSVK_25} This is an illustration of the abstract object by which we model the class of equivalence of dense molecular clouds.}
\end{figure}

In this section we shortly introduce the model for our dense molecular cloud. First of all we recall the notion "ensemble of MCs" or alternatively "MC class of equivalence" \cite{DVK_17}. One class of equivalence is a set of MCs with the same mass density PDF, the same effective size, and the same density at the outer edge. In this work we regard only irrotational dense molecular clouds. Every class of equivalence has one abstract object - "ensemble abstract representative member", which is an irrotational spherically symmetric and isotropic dense cloud (we realise that the real objects exhibit deviations from spherical form, isotropy and also may have angular momentum). At the centre of the abstract object is located a very small and dense spherical volume (or core), which is a homogeneous ball (see FIG. \ref{fig:Fig_1_DSVK_25}. for illustration).
We suppose that the size and mass of the core are much less than the size and mass of the cloud: $l_{\rm c} \gg l_0$, and $M_{\rm c} \gg M_0$ (where $l_{\rm c}$ and $M_{\rm c}$ are size and mass of the cloud, and $l_0$ and $M_0$ are size and mass of the core, accordingly). We consider accretion of gas through the cloud's outer boundary. The material passes radially through all the cloud's scales and flows into the core, but does not change its size and mass for several dynamical times (the time range within which we study the system). We regard the gas to be isothermal, at constant temperature $T\sim 10-30 {\rm K}$, typical for MCs. Accordingly, the sound speed $c_{\rm s}$ is constant. The cloud is also turbulent, as the turbulence is fully saturated and all the cloud's scales (radii) belong to the inertial range. In every spherical shell of the cloud this turbulence is locally homogeneous and isotropic. We suppose that for the mass density field there exists a density profile, which reads:
\begin{equation}
	\label{mass-density profile-def}
	\varrho = \ell^{-p}~~,~~p=p(\ell,\tau)~,
\end{equation}
where $p=p(\ell,\tau)$ is the exponent of the profile, depending on both: radius $\ell_0 \leq \ell \leq 1$ and time $\tau$. From now on we will use dimensionless quantities for scales (radii), density and velocity of the fluid, and time, as follows: 
\begin{equation}
	\label{dimless quantities}
	\ell\equiv l/l_{\rm c},~\varrho\equiv\rho/\rho_{\rm c},~v\equiv u/c_{\rm s},~\tau\equiv tc_{\rm s}/l_{\rm c}~,
\end{equation} 
where $\rho_{\rm c}$ is the outer edge density. We assume, as well, that the cloud is immersed in a very large medium, whose matter is distributed radially symmetric in respect to the cloud's centre (in particular, the medium may be homogeneous), hence this medium does not exert gravitational force on a fluid element residing in the cloud's interior. For more details the reader may see Donkov \& Stefanov \cite{DS_18,DS_19}, where the model was originally set up.

\section{Evolution of the density profile}
\label{sec-Evolution of the density profile}

\subsection{Equations of the medium}
\label{subsec-equations of the medium}

We start from the equations of the medium in their dimensional form. They are:

- The system of compressible Euler equations

\begin{equation}
	\label{equ_N-St}
	\frac{\partial\vec{u}}{\partial t} + \vec{u}\cdotp\nabla\vec{u} = - \frac{1}{\rho}\nabla P_{\rm th} - \nabla\varphi
\end{equation}

\begin{equation}
	\label{equ_cont}
	\frac{\partial\rho}{\partial t} + \nabla\cdotp(\rho\vec{u})=0
\end{equation}

- The equation of state of the gas

\begin{equation}
	\label{equ_id-gas}
	P_{\rm th}=c_{\rm s}^2 \rho
\end{equation}

- The Poisson equation for the gravitational potential

\begin{equation}
	\label{equ_Pois}
	\Delta\varphi=4\pi G \rho~.
\end{equation}

As it can be seen, in the equation (\ref{equ_N-St}), which is, indeed, the equation of motion of a fluid element, the dissipative terms have been neglected. This is justified due to the statistical equilibrium which is characteristic for the inertial interval of scales. The equation of continuity (\ref{equ_cont}) is actually a differential form of the law of conservation of mass.

The equation of state of the ideal gas (\ref{equ_id-gas}) reflects the assumption that the medium is isothermal. The Poisson equation (\ref{equ_Pois}) determines the gravitational potential produced by the given density distribution.

Other factors, such as the angular momentum of the core, the magnetic fields, the feedback from the newborn stars, have been deliberately neglected, so that the model is not too cumbersome for this attempt of obtaining the equations governing the evolution of mass density profile.

We should note here, that without accretion, the set of equations (\ref{equ_N-St}) - (\ref{equ_Pois}) have been solved many years ago, leading to the Lane-Emden solution for isothermal self-gravitating spheres. May be, the most prominent one is the Bonnor-Ebert sphere \cite{Bonnor_56,Ebert_57}, which is the Lane-Emden solution without singularity in the centre (i.e. finite central density) and with confining pressure at the outer cloud's boundary. Although, in the outer layers Bonnor-Ebert sphere displays density profile $\varrho\sim\ell^{-2}$, coinciding with density profile in our model (far from the core), under the condition of steady state, the two models are different. Their sphere is in hydrostatic equilibrium, while our object is characterized by a stationary accretion flow through the cloud's boundary.

\subsection{Equations governing the evolution of the density profile}
\label{subsec-equations governing the evolution of the density profile}

To obtain the equations governing the evolution of the mass density profile we first rewrite the equations of the medium in terms of dimensionless variables $\ell,~\varrho,~v,$ and $\tau$, which were introduced in Section \ref{sec-The model}. Then we note that the velocity field is a vector sum of two components: $\vec{v} = \vec{v}_{\rm a} + \vec{v}_{\rm t}$, which account for the radial accretion ($\vec{v}_{\rm a}$) of gas to the centre and for the chaotic turbulent motions ($\vec{v}_{\rm t}$), accordingly. In the inertial range, which we assume to be in statistical equilibrium, and after ensemble averaging of the Euler system with respect to the microstates of the turbulence, we find that the turbulent component can be neglected in comparison to the thermal pressure term. This is due to the homogenous and locally isotropic nature of turbulence and trans- or subsonic Mach number of the turbulent flow for the whole cloud. Short formal analysis shows that the l.h.s. of equation (\ref{equ_N-St}) can be rewritten as follows:

$$\frac{\partial \vec{v}}{\partial \tau} + \vec{v}\cdot\nabla\vec{v} = \frac{\partial (\vec{v}_{\rm a} + \vec{v}_{\rm t})}{\partial \tau} + (\vec{v}_{\rm a} + \vec{v}_{\rm t})\cdot\nabla\ (\vec{v}_{\rm a} + \vec{v}_{\rm t}) = $$
$$ \frac{\partial \vec{v}_{\rm a}}{\partial \tau} + \frac{\partial \vec{v}_{\rm t}}{\partial \tau} + \vec{v}_{\rm a}\cdot\nabla\ \vec{v}_{\rm a} + \vec{v}_{\rm a}\cdot\nabla\vec{v}_{\rm t} + \vec{v}_{\rm t}\cdot\nabla\ \vec{v}_{\rm a} + \vec{v}_{\rm t}\cdot\nabla\vec{v}_{\rm t}$$
After averaging, due to the vanishing of the turbulent velocity field: $\langle \vec{v}_{\rm t} \rangle_{\rm e} = 0$, which is assumed to be locally purely chaotic, all the terms of first order in regard to $\vec{v}_{\rm t}$ vanish. Therefore one has:

$$\frac{\partial \vec{v}_{\rm a}}{\partial \tau} + \vec{v}_{\rm a}\cdot\nabla\ \vec{v}_{\rm a} + \langle \vec{v}_{\rm t}\cdot\nabla\vec{v}_{\rm t} \rangle_{\rm e} = \frac{{\rm d} \vec{v}_{\rm a}}{{\rm d} \tau} + \langle \vec{v}_{\rm t}\cdot\nabla\vec{v}_{\rm t} \rangle_{\rm e} ~,$$
The last term above, if one accounts for the isotropy of the turbulent velocity field, presents the gradient of turbulent pressure: $\langle \vec{v}_{\rm t}\cdot\nabla\vec{v}_{\rm t} \rangle_{\rm e} = \nabla \langle v_{\rm t}^2/2 \rangle_{\rm e}$. According to our model presented in \cite{DS_19}, the turbulent pressure within the inertial range scales as follows: $\langle v_{\rm t}^2 \rangle_{\rm e} = T_0\ell^{2\alpha} $, where the scaling exponent resides in the range $0\leq \alpha \leq 1$ (see \cite{Elme_Scalo_04,Mac Low_Klessen_2004,Federrath_ea_10,HF_12}) and does not depend on the scale and time. The dimensionless coefficient $T_0$ is the ratio of the turbulent energy per unit mass to the thermal energy per unit mass, at the cloud's edge, where $\ell=1$ (this is, indeed, the squared turbulent Mach number for the cloud as a whole). Hence $\nabla \langle  v_{\rm t}^2/2 \rangle_{\rm e} = (\alpha /2) T_0 \ell^{2\alpha-1} \hat{\ell}$, where $\hat{\ell}$ is the unit vector directed radially outwards. The turbulent pressure gradient term appears in the general equations obtained bellow (see equation \ref{1-equ}), but is subdominant compared to the thermal pressure gradient term, under the assumption $T_0 \lesssim 1$ (sub- or trans-sonic turbulent flow), in the solutions we obtain under the steady state flow condition (see section \ref{subsec-analytical solution in the case of steady-state flow}). We note that the assumption for sub- or trans-sonic turbulent flow are valid throughout this paper. In order to complete the consideration of averaging of the equations of the medium we have to say that the r.h.s. of equation (\ref{equ_N-St}) is averaged according to the model, because $P_{\rm th}$ and $\varphi$ are the mean thermal pressure and gravity potential of our cloud, respectively. In a similar manner, the l.h.s. of the equation (\ref{equ_cont}) can be averaged under the assumption for the mean density field $\varrho$ characterizing our abstract cloud:

$$\bigg\langle \frac{\partial\varrho}{\partial \tau} + \nabla\cdotp(\varrho\vec{v}) \bigg\rangle_{\rm e} = \frac{\partial\varrho}{\partial \tau} + \nabla\cdotp(\varrho\langle \vec{v}_{\rm a} + \vec{v}_{\rm t} \rangle_{\rm e}) = \frac{\partial\varrho}{\partial \tau} + \nabla\cdotp(\varrho\vec{v}_{\rm a})~.$$
We stress, also, that all calculations are done in a Eulerian spherical coordinate system with origin located in the centre of the cloud. All functions depend only on $\ell-$coordinate and time, according to the model. Let us continue with the formula for the dimensionless gravitational force per unit mass, acting on a fluid element: $$- \nabla\phi = -(3/2)G_0 \frac{1}{\ell^2} \bigg[ \int_{\ell_0}^{\ell} \ell'^{2-p(\ell',\tau)} {\rm d}\ell' \bigg] \hat{\ell} - \frac{G_1}{2\ell^2} \hat{\ell}~,$$
where the first term accounts for the shells of the cloud located between the fluid element and the central volume, and the second term is the gravitational pull on the fluid element from this central volume, respectively (note that the unit vector $\hat{\ell}$ is directed radially outwards and the dimensionless gravitational potential reads: $\phi\equiv\varphi/c_{\rm s}^2$). The formula for $- \nabla\phi$, indeed, comes from the solution $\phi$ of the Poisson equation for our model. The dimensionless coefficients $G_0 \equiv (4/3)\pi G \rho_{\rm c} l_{\rm c}^2 / (c_{\rm s}^2/2)$ and $G_1 \equiv GM_0/l_{\rm c}(c_{\rm s}^2/2)$ have the following physical  interpretation, respectively: the ratio of the gravitational energy of the entire cloud (without the core)  and its thermal energy per unit mass, and the ratio of the gravitational energy per unit mass of the core and its thermal energy per unit mass. Both of them are evaluated at the outer boundary of the cloud. We note also that the second term on the l.h.s. of equation (\ref{equ_cont}) reads $(1/\ell^2)\partial(-\ell^2\varrho v_{\rm a})/\partial\ell$ (after averaging and due to the spherical symmetry). Some additional calculations, which might be useful:

- the pressure gradient term in equation (\ref{equ_N-St}):

$$-\frac{1}{\varrho}\nabla\varrho = -\nabla\ln\varrho = $$
$$ \frac{{\rm d}}{{\rm d}\ell}(-p\ln\ell)(-\hat{\ell}) = \bigg( - (\ln\ell) \frac{\partial p(\ell,\tau)}{\partial \ell} - \frac{p(\ell,\tau)}{\ell} \bigg)(-\hat{\ell})~;$$

- the two partial derivatives in equation (\ref{equ_cont}):

$$\frac{\partial\varrho}{\partial \tau} = -(\ln\ell)\ell^{-p}\frac{\partial p(\ell,\tau)}{\partial \tau}~;$$
and

$$-\frac{1}{\ell^2}\frac{\partial}{\partial\ell}(\ell^2\varrho v_{\rm a}) = -\frac{1}{\ell^2}\frac{\partial}{\partial\ell}(\ell^{2-p} v_{\rm a}) = $$
$$ -\ell^{-p} \bigg[ \frac{\partial v_{\rm a}}{\partial \ell} + \bigg( \frac{2-p(\ell,\tau)}{\ell} - (\ln\ell) \frac{\partial p(\ell,\tau)}{\partial \ell} \bigg) v_{\rm a} \bigg]~.$$
Finally, one gets the system of two equations:

\begin{eqnarray}
	\label{1-equ}
	\frac{{\rm d}v_{\rm a}(\ell,\tau)}{{\rm d}\tau} = \frac{3}{2} G_0 \frac{1}{\ell^2} \int_{\ell_0}^{\ell} \ell'^{2-p(\ell',\tau)} {\rm d}\ell' + \frac{G_1}{2\ell^2} \nonumber\\ 
	- (\ln\ell) \frac{\partial p(\ell,\tau)}{\partial \ell} - \frac{p(\ell,\tau)}{\ell} - \frac{\alpha}{2} T_0 \ell^{2\alpha-1} ~,
\end{eqnarray}
and

\begin{eqnarray}
	\label{2-equ}
	- (\ln\ell) \frac{\partial p(\ell,\tau)}{\partial \tau} = \frac{\partial v_{\rm a}(\ell,\tau)}{\partial \ell} + \nonumber\\
	\bigg( \frac{2-p(\ell,\tau)}{\ell} - (\ln\ell) \frac{\partial p(\ell,\tau)}{\partial \ell} \bigg) v_{\rm a}(\ell,\tau)~.
\end{eqnarray}
This first-order partial differential equation system governs the evolution of the function $p=p(\ell,\tau)$: the exponent of mass density profile. This system contains two unknown fields: $p(\ell,\tau)$ and $v_{\rm a}(\ell,\tau)$. The first equation (equation \ref{1-equ}) is integro-differential and non-linear due to the first term on the r.h.s. The second equation (equation \ref{2-equ}) is also non-linear. It is hardly probable that the system has an analytical solution. It might be solved numerically, with a careful choice of initial and boundary conditions.

\subsection{Analytical solution in the case of steady-state flow}
\label{subsec-analytical solution in the case of steady-state flow}

Despite its complexity this system can give us analytical expressions in the case of steady state flow, when the unknown fields do not depend explicitly on time $\tau$, i.e. $p=p(\ell)$ and $v_{\rm a}=v_{\rm a}(\ell)$. In this case $\partial p/\partial \tau = 0$, $\partial p/\partial \ell \equiv {\rm d} p/ {\rm d} \ell$, $\partial v_{\rm a}/\partial \ell \equiv {\rm d} v_{\rm a}/ {\rm d} \ell$, and ${\rm d} v_{\rm a}/ {\rm d} \tau = ({\rm d} v_{\rm a}/ {\rm d} \ell)\times ({\rm d} \ell/ {\rm d} \tau) = - v_{\rm a} ({\rm d} v_{\rm a}/ {\rm d} \ell)$ (note that the accretion velocity is directed inwards). Also, the observations and simulations show that, in the steady state case, the density profile has an exponent, which is a constant in respect to the cloud scales, hence in our model we suppose that: $\partial p/\partial \ell = 0$ (i.e. $p={\rm const.}$). Therefore the equations become much simpler, as follows:

\begin{eqnarray}
	\label{1-equ-steady-state}
	(-v_{\rm a})\frac{{\rm d}v_{\rm a}}{{\rm d}\ell} = \frac{3}{2} G_0 \frac{\ell^{1-p} ( 1 - [\ell_0/\ell]^{3-p} )}{3-p} + \frac{G_1}{2\ell^2} \nonumber\\
	- \frac{p}{\ell} - \frac{\alpha}{2} T_0 \ell^{2\alpha-1}~,
\end{eqnarray}
and

\begin{equation}
	\label{2-equ-steady-state}
	0 = \frac{{\rm d} v_{\rm a}}{{\rm d} \ell} + \frac{2-p}{\ell} v_{\rm a}~.
\end{equation}
The second equation (equation \ref{2-equ-steady-state}) can be solved solely for the accretion velocity field, and one gets, $v_{\rm a} = \sqrt{A_0} \ell^{p-2}$, where $\sqrt{A_0}$ is a coefficient corresponding to the boundary condition at the outer edge of the cloud ($\ell=1$), i.e. $\sqrt{A_0}=v_{\rm a,c}$. This is the velocity of the fluid, when it crosses the cloud's boundary. $A_0$ can be also interpreted as the ratio of the accretion and the thermal energy, at the cloud's boundary \cite{DS_19}. The equation (\ref{1-equ-steady-state}) is the equation of motion of a fluid element. Solving it far from the cloud's centre ($\ell\gg\ell_0$) and assuming that $3G_0 \gg G_1$, one can neglect the gravity of the central volume, presented by the term: $G_1/2\ell^2$, in respect to the gravity of cloud's layers bellow the fluid element, and also simplify the expression in the parentheses on the r.h.s.: $1 - [\ell_0/\ell]^{3-p} \approx 1$, as far as according to observations and simulations $p<3$, typically. Then substituting the expression for the accretion velocity in equation (\ref{1-equ-steady-state}), one obtains:

\begin{equation}
	\label{equ-p-steady-state}
	0 = (2p-4)\frac{A_0 \ell^{2p-5}}{2} + \frac{3}{2} G_0 \frac{\ell^{1-p}}{3-p} - \frac{p}{\ell} - \frac{\alpha}{2} T_0 \ell^{2\alpha-1}~.
\end{equation}
The latter equation (which is an equation for $p$) can be solved approximately through the method based on the balance of the leading order terms \cite{DS_19,Zhivkov_99,RHB_2006}. This method consists in comparing every two terms in the equation and obtaining equations for $p$. The solutions one gets must be substituted in (\ref{equ-p-steady-state}) and only the leading order terms in respect to the powers of $\ell$ must be taken into account (note that $\ell<1$, and hence the leading order terms are those of the smallest power). One has a solution for $p$, if the leading order terms are more than one and they balance the equation. Using this technics one obtains the only possible steady-state solution, far from the cloud's centre: $p=2$ (the leading order terms are $\sim\ell^{-1}$). The mass density profile reads: $\varrho=\ell^{-2}$, and equation (\ref{equ-p-steady-state}) gives: $(3/2)G_0\ell^{-1} - 2\ell^{-1}=0~\Rightarrow~G_0=4/3$, whose physical meaning is that steady state establishes when gravity of the layers bellow a fluid element is balanced by the gas pressure gradients (of the same layers). (Recall that the turbulent pressure term is subdominant compared to the thermal pressure term for all $0< \alpha \leq 1$ and vanishes for $\alpha=0$. So it does not affect the solution.) In that case accretion velocity field is constant: $v_{\rm a} = \sqrt{A_0} = v_{\rm a,c}$. Here, we should note that, if $v_{\rm a,c}=0$, then $v_{\rm a}\equiv0$ and hence one has a static equilibrium in the layers far from the core. Then the solution for the profile is again $p=2$, which stems from equation (\ref{equ-p-steady-state}). This equation gives, also, that the balance is between the gravity of the layers under the fluid element from the one side and their pressure gradient from the other. But the latter equilibrium is hydrostatic and we obtain the Bonnor-Ebert profile \cite{Bonnor_56,Ebert_57} far from the centre. This solution for $p$, far from the cloud's core, and the corresponding solutions for the two unknown fields: $\varrho=\ell^{-2}$ and $v_{\rm a} = v_{\rm a,c}$, coincide completely with the solution obtained by using the equation of energy conservation of a fluid element \cite{DS_19}. It follows that: $A_0 - 3G_0 = E_0={\rm const.}$, where $E_0$ is the energy of a fluid element through the steady state condition. So, if one can asses the coefficient $A_0$, then will obtain the value of $E_0$, which is a measure for the stability of cloud's dynamical state. It is worth to note, that the solution $p=2$, far from the centre, in the context of the energetic approach used in \cite{DS_19} exists for all values of $A_0>0$. The latter is true due to the total energy of the fluid element $E_0$ balances the equation in the sense, that: if $A_0\ll3G_0$, then the energy balance is $-3G_0=E_0$ and the energy of the fluid element is negative; on the contrary, if $A_0\gg3G_0$, then $A_0=E_0$ and the energy of the fluid element is positive. In both cases the solution for $p$ exists and is $p=2$ \cite{DS_19}. So, if the flow is trans- or subsonic, then $A_0\leq1$, and $E_0\leq 1-4=-3<0$, hence the outer layers of the core are in a stable dynamical state. In the case of moderately supersonic flow with Mach number $M\approx2$ (note that $A_0=M^2$ for the accretion flow), one has $A_0\approx4$ and $E_0\approx0$, which supposes a marginally bound dynamical state for the outer layers. If $M>2$, then $A_0>4$, and therefore $E_0>0$ meaning that the outer layers are dynamically unstable. The presented stability analysis coincides with the one obtained by using the virial theorem \cite{DSVK_22}.

Also, one can obtain, from equations (\ref{1-equ-steady-state}) and (\ref{2-equ-steady-state}), the solution near to the cloud's core, where $\ell \approx \ell_0$. Then in the equation (\ref{1-equ-steady-state}) the first term on the r.h.s. vanishes, because $1 - [\ell_0/\ell]^{3-p} \approx 0$. The gravity of the core dominates over the gravity of the very thin layers under the fluid element. Hence the equation of motion reads:

\begin{equation}
	\label{equ-p-near the core}
	0 = (2p-4)\frac{A_0 \ell^{2p-5}}{2} + \frac{G_1}{2\ell^2} - \frac{p}{\ell} - \frac{\alpha}{2} T_0 \ell^{2\alpha-1}~.
\end{equation}
Using the same methodology for obtaining solutions for $p$ \cite{DS_19}, we arrive at the only possible solution: $p=3/2$, where the equation for the coefficients is: $-A_0\ell^{-2}/2 + G_1\ell^{-2}/2 = 0~\Rightarrow~A_0=G_1$ (the leading order terms are $\sim\ell^{-2}$). The density and velocity fields are as follows: $\varrho=\ell^{-3/2}$ and $v_{\rm a} = \sqrt{A_0} \ell^{-1/2}$. The obtained solution, physically, is simply a free fall of a fluid element under the gravitational pull of the core. In regard to the flow stability, this state is marginally stable, as the energy of the fluid element is zero. This results completely coincides with the one obtained in \cite{DS_19}, for the case near to the cloud's core, by using the equation of energy conservation of a fluid element.

It is important to note here, that both solutions: far from the core, and near to it, are consistent, in a sense that they can be ascribed to the same stage of evolution of the cloud, only if $3G_0 \gg A_0>0$ (i.e. $M<1$ - subsonic flow), due to the condition $3G_0 \gg G_1$, which allows the existence of the solution $p=2$, far from the centre. Since, the solution near to the core exists for any $A_0>0$, then for accretion flows with $M>1$ the existence of solution $p=2$ far from the centre is problematic. Hence, if the two solutions exist simultaneously, in the same epoch, then $G_1\ll 3G_0$, $A_0\ll 3G_0$, therefore $E_0<0$ and the outer layers are in a stable dynamical state.

\subsection{The steady state solution is an attractor}
\label{subsec-the steady-state solution is an attractor}

In this section we show that the slope of steady state solution $p=2$ for the density profile, far from the central volume, is an attractor. We conduct our considerations under two assumptions, which are as follows: the first one is the scaling exponent of the density field does not depend on the scale $\ell$, but only on time $\tau$, so $p=p(\tau)$, and hence $\partial p/\partial \ell = 0$, $\partial p/\partial \tau \equiv {\rm d} p/ {\rm d} \tau$; the second one is that the velocity field also follows a scaling law: $v_{\rm a}=a\ell^{\beta(\tau)}$, where $a$ is a constant determined by the boundary condition (at the outer edge of the cloud) and the exponent $\beta(\tau)$ depends only on time, hence $\partial \beta/\partial \ell = 0$. Then the equations (\ref{1-equ}) and (\ref{2-equ}) read:

\begin{equation}
	\label{equ-1-atr}
	\frac{{\rm d}v_{\rm a}}{{\rm d}\tau} = \frac{3}{2} G_0 \frac{\ell^{1-p}}{3-p} - \frac{p}{\ell} - \frac{\alpha}{2} T_0 \ell^{2\alpha-1}~.
\end{equation}
and

\begin{equation}
	\label{equ-2-atr}
	- (\ln\ell) \frac{{\rm d} p}{{\rm d} \tau} = \frac{\partial v_{\rm a}}{\partial \ell} + \frac{2-p}{\ell}v_{\rm a} = \frac{2-p+\beta}{\ell}v_{\rm a}~.
\end{equation}
These two equations give us that if $p>2$, then the first term (gravity of the layers under the fluid element) on the r.h.s. of equation (\ref{equ-1-atr}) is of leading order in regard to the powers of $\ell$ and hence the acceleration of the fluid element is positive, therefore the velocity $v_{\rm a}$ will increase during the motion (of the fluid element) to the centre. The latter means that the exponent $\beta<0$ must be negative and the r.h.s. of equation (\ref{equ-2-atr}) is also negative, hence the derivative ${\rm d} p/{\rm d} \tau < 0$ is negative, too (note that $-\ln\ell>0$). Therefore the function $p(\tau)$ decreases. On the contrary, if $p<2$, then the second term (gas pressure gradient) on the r.h.s. of equation (\ref{equ-1-atr}) is of leading order, hence the acceleration of the fluid element is negative, i.e. its velocity will decreases to the centre. This assumes that $\beta>0$ is positive, and the r.h.s. of equation (\ref{equ-2-atr}) is positive, too. Hence the derivative ${\rm d} p/{\rm d} \tau > 0$ is also positive. Therefore the function $p(\tau)$ increases. The conclusion is that the steady state value $p=2$ is an attractor. We note that all the considerations are conducted through the assumption that $1<p<3$ according to many simulations and observations \cite{Klessen_2000,KNW_11,Girichidis_ea_14,Veltchev_ea_19,Khullar_ea_21,Schneider_ea_2022}.

\section{Discussion}
\label{sec-discussion}

Our results are similar to previous works dedicated to the study of density and velocity profiles of self-gravitating isothermal spheres. The solution of the hydrostatic Lane-Emden equations for Bonnor-Ebert spheres \cite{Bonnor_56,Ebert_57}, far from the centre, displays density profile $\varrho\sim\ell^{-2}$, coinciding with our solution far from the core. The same profile, in the outer cloud's layers, is reported by Shu \cite{Shu_77} (hydrostatic equilibrium) and Larson-Penston \cite{Larson_69,Penston_69a} for collapsing initially homogeneous super-Jeans spheres. Also, our solution near to the core (for both density and velocity field) coincides with Shu's and Larson-Penston's solution for the cloud's interior after the appearance of the central point mass (i.e. the protostar). These works have been generalized by Whitworth \& Summers \cite{Whitworth_Summers_85}, who investigated the whole parametric space of the so called "similarity solutions" for self-gravitating isothermal spheres. They obtained three asymptotic solutions and two of them, for $t=0$, when the point mass occurs due to the supersonic compression wave, and $t\rightarrow +\infty$, correspond to our solutions (for both density and velocity field) far from the core and near to the core, respectively. The latter authors also found that the accretion velocity must be supersonic in these two cases, which is outcome of the initial and boundary conditions of the problem (the collapse of the sphere starts with uniform density and zero velocity). Our model allows sub- and trans-sonic velocities and, in particular, in these cases the outer layers of the cloud are in a stable dynamical state. This is possible, if the larger medium (giant molecular cloud), where our dense cloud is submerged, has not started to collapse as a homogeneous sphere at rest. The contemporary point of view \cite{KG_16,BP_ea_20} about the structure and dynamics of MCs considers they have complicated filamentary structure and our dense clouds are located at intersection points of several filaments, which feed the clouds with gas. If MCs are collapsing at larger scales, this collapse likely is hierarchical and chaotic \cite{VS_ea_19,BP_ea_20}. These complicated structure and dynamics can substantially decrease the accretion velocities at the scale of dense structures, where star formation takes place. In a resent paper, Wang et al. \cite{Wang_ea_24} observe many dense clouds ("cores" in their terminology) in which the non-thermal velocity dispersions are mostly sub- or trans-sonic (although, there are "cores" displaying moderately supersonic non-thermal velocity dispersions), which is consistent with our conclusion for stability of the outer layers. Indeed, the latter authors called this "sub- or trans-sonic turbulence", but this is probably the accretion flow caused by the hierarchical and chaotic collapse of the medium from larger down to smaller scales \cite{VS_ea_19,BP_ea_20}.

In the second application of the general equations (\ref{1-equ} and \ref{2-equ}), we show, in section \ref{subsec-the steady-state solution is an attractor}, that the steady state exponent of the density scaling law $p=2$ is an attractor, and all slopes (in the range: $1<p<3$), which depend on time, converge to the steady state solution. This important result we get under the assumption that $p$ depends only on time (i.e. $\partial p/\partial \ell = 0$) and the accretion velocity field scales as follows: $v_{\rm a}=a\ell^{\beta(\tau)}$, where $\partial \beta/\partial \ell = 0$. The latter assumptions are justified by many observations and simulations \cite{Elme_Scalo_04,Federrath_ea_10,Girichidis_ea_14,HF_12,Gomez_ea_2021,KG_16,KNW_11,Klessen_2000,Larson_69,Mueller_ea_02,Evans_03,Pad_ea_06,N-R_ea_15,Penston_69a,Veltchev_ea_19,Wang_ea_24,Wyr_ea_12,Wyr_ea_16,Zhang-Li_17,Csengeri_ea_17,Schneider_ea_2022}. Also, we use again method of leading order terms in respect to the powers of $\ell$. These simplifications mean that the result is approximate, but nevertheless the main conclusion is that the steady state solution is the most important one and the slope $p\approx2$ should be observable in dense clouds (please note that in observational papers these objects are called "dense cores").

In a recent paper, Gomes, Vazquez-Semadeni \& Palau \cite{Gomez_ea_2021} investigated the evolution of density profile of a prestellar core (dense cloud in our terminology), submerged in a homogeneous medium and accreting gas from it. They obtained (under the assumption that $p$ does not depend on radius) an equation (their equation 13) for the evolution of density profile exponent $p(t)$ and concluded that the value $p=2$ is a stationary solution and it is an attractor for the time dependent solutions. The difference in respect to our approach is that they derived the equation for $p(t)$ applying only the continuity equation and assuming that the accretion velocity field is determined only by self-gravity, i.e. the motion of a fluid element is a free falling, while we regard the equation of motion with two forces: self-gravity and gas pressure gradient, which act opposite to each other, which is absent in their construction.

Finally, we note that the presented work is based on several simplifications, which differ from the real objects. In particular, these are spherical symmetry of the mass distribution and radial accretion flow, although real objects are very close to these conditions. The regarded model present an irrotational cloud, although the real clouds may have non-negligible angular momentum. Also, the proof that $p=2$ is an attractor was made under the assumption that the exponents of the scaling laws for density and velocity fields do not depend on scale, which is true only in the extreme cases: far from the core or near to it. Nevertheless, we regard our work as a next step in exploring the key problem of evolution of the mass density profile of dense molecular clouds.

\section{Summary}
\label{sec-summary}

In this paper we investigate the equations that govern the evolution of the mass density profile of dense molecular clouds. Starting from spherical symmetry and isotropy of the matter and gas flows we arrived at equations (\ref{1-equ}) and (\ref{2-equ}), which govern the dynamics of the unknown fields: $p(\ell,\tau)$ - the exponent of the mass density profile, and $v_{\rm a}(\ell,\tau)$ - the accretion velocity of the gas. The system is of first-order and the partial differential equations are non-linear in regard to both unknown fields. Moreover, the equation (\ref{1-equ}) is integro-differential. Despite the complicated nature of the equations, they lead to analytical solutions for the unknown fields in case of a steady state flow. We could simplify the original system and obtained explicit solutions in two regimes: far from the cloud's central volume (core) and near to it. In the former regime we obtain that: $\varrho=\ell^{-2}$ and $v_{\rm a}={\rm const.}$, while in the latter regime we get: $\varrho=\ell^{-3/2}$ and $v_{\rm a}=\sqrt{A_0}\ell^{-1/2}$. Moreover, we arrived at equations for dimensionless coefficients, which are energies of a fluid element per unit mass. Far from the central volume we have: $G_0=4/3$, and near to it the equation is: $A_0=G_1$. All these results (with exception for the equation of energy balance, far from the centre), for a steady-state flow, completely coincide with those obtained in \cite{DS_19} by applying the equation of energy conservation of a fluid element. In the present work we reached the same by using the dynamical equations of the model. It is worth to note, that combining the two different equations of energy balance, far from the centre, and regarding different Mach numbers for the accretion velocity at the cloud's boundary, one can conclude when the outer layers of the cloud are in a stable dynamical state, and when they are not. We should note, that both solutions: near to the core, and far from it, are consistent only if the accretion flow is subsonic, and in that case the outer layers are in a stable dynamical state. Also, we may speculate that at earlier stages of cloud's evolution the cloud's core is not too massive, i.e. $3G_0\gg G_1$ and then the solution $p=2$ describes the steady state flow in the outer layers of the cloud, without any restriction on the accretion velocity (i.e. for all the values $A_0>0$). We note that the method of obtaining  the solutions in steady state case is approximate, based on the finding of leading order terms (which balance the equation) in respect to the exponents of scale $\ell$. The same method we used in section \ref{subsec-the steady-state solution is an attractor} to show that the steady state solution $p=2$, far from the cloud's centre, is an attractor and all slopes (in the range: $1<p<3$), which depend on time, converge to it. This is the second application of the general equations (\ref{1-equ} and \ref{2-equ}). It indicates that the steady state solution, far from the centre, is the most important one and should be observable in dense molecular clouds.

\begin{acknowledgments}
	S.D. acknowledges funding by the Ministry of Education and Science of Bulgaria (support for the Bulgarian National Roadmap for Research Infrastructure) and for the support by DFG under grant KL 1358/20-3. T.V. acknowledges support by the DFG under grant KL 1358/20-3 and additional funding from the Ministry of Education and Science of the Republic of Bulgaria, National RI Roadmap Project DO1-326/4.12.2023. R.S.K. acknowledges financial support from the European Research Council via the  Synergy Grant ``ECOGAL'' (project ID 855130),  from the German Excellence Strategy via the Heidelberg Excellence Cluster  ``STRUCTURES'' (EXC 2181 - 390900948), and from the German Ministry for Economic Affairs and Climate Action in project ``MAINN'' (funding ID 50OO2206). R.S.K. also thanks the 2024/25 Class of Radcliffe Fellows for their company and for highly interesting and stimulating discussions.
\end{acknowledgments}


\begin{thebibliography}{40}
	
\bibitem[\protect\citeauthoryear{Elmegreen \& Scalo}{2004}]{Elme_Scalo_04}
B. Elmegreen, J. Scalo, Ann. Rev. Astron. \& Astrophys., {\bf 42}, 211E (2004)
\bibitem[\protect\citeauthoryear{Hennebelle \& Falgarone}{2012}]{HF_12}
P. Hennebelle, E. Falgarone, Ann. Rev. Astron. \& Astrophys., {\bf 20}, 55H (2012)
\bibitem[\protect\citeauthoryear{Klessen \& Glover}{2016}]{KG_16}
R. S. Klessen, S. C. O. Glover, Star Formation in Galaxy Evolution: Connecting Numerical Models to Reality, Saas-Fee Advanced Course, {\bf 43}, ISBN 978-3-662-47889-9. Springer-Verlag Berlin Heidelberg, p.85 (2016)
\bibitem[\protect\citeauthoryear{Ferriere}{2001}]{Ferriere_01}
K. M. Ferriere, Reviews of Modern Physics, {\bf 73}, 1031 (2001)
\bibitem[\protect\citeauthoryear{Wang et al.}{2010}]{Wang_ea_2010}
P. Wang, Z.-Y. Li, T. Abel, F. Nakamura, The Astrophys. Journal, {\bf 709}, 27 (2010)
\bibitem[\protect\citeauthoryear{Federrath}{2015}]{Federrath_15}
C. Federrath, Mon. Not. R. Astron. Soc., {\bf 450}, 4035 (2015)
\bibitem[\protect\citeauthoryear{Mathew \& Federrath}{2021}]{Mathew_Federrath_21}
S. S. Mathew, C. Federrath, Mon. Not. R. Astron. Soc., {\bf 507}, 2448 (2021)
\bibitem[Lombardi, Alves \& Lada (2010)]{LAL_2010}
Lombardi, M., Alves, J. \& Lada C., A\&A, {\bf 519}, 7, (2010)
\bibitem[\protect\citeauthoryear{V\'azquez-Semadeni}{1994}]{VS_94}
E. V\'azquez-Semadeni, The Astrophys. Journal, {\bf 423}, 681 (1994)
\bibitem[\protect\citeauthoryear{Padoan et al.}{2006}]{Pad_ea_06}
P. Padoan, M. Juvela, A. Kritsuk, M. Norman, The Astrophys. Journal, {\bf 653}, 125 (2006)
\bibitem[\protect\citeauthoryear{Federrath et al.}{2010}]{Federrath_ea_10}
C. Federrath, J. Roman-Duval, R. S. Klessen, W. Schmidt, M.-M. Mac Low, Astron. \& Astrophys., {\bf 512}, 81 (2010)
\bibitem[\protect\citeauthoryear{Molina et al.}{2012}]{Molina_ea_12}
F. Molina, S. C. O. Glover, C. Federrath, R. S. Klessen, Mon. Not. R. Astron. Soc., {\bf 423}, 2680 (2012)
\bibitem[Klessen (2000)]{Klessen_2000}
R. S. Klessen, The Astrophys. Journal, {\bf 535}, 869 (2000)
\bibitem[\protect\citeauthoryear{Kritsuk, Norman \& Wagner}{2011}]{KNW_11}
A. Kritsuk, M. Norman, R. Wagner, The Astrophys. Journal, {\bf 727}, 20 (2011)
\bibitem[\protect\citeauthoryear{Girichidis et al.}{2014}]{Girichidis_ea_14}
P. Girichidis, L. Konstandin, A. Whitworth, R. S. Klessen, The Astrophys. Journal, {\bf 781}, 91 (2014)
\bibitem[\protect\citeauthoryear{Veltchev et al.}{2019}]{Veltchev_ea_19}
T. V. Veltchev, P. Girichidis, S. Donkov, N. Schneider, O. Stanchev, L. Marinkova, D. Seifried, R. S. Klessen, Mon. Not. R. Astron. Soc., {\bf 489}, 788 (2019)
\bibitem[\protect\citeauthoryear{Khullar et al.}{2021}]{Khullar_ea_21}
S. Khullar, C. Federrath, M. R. Krumholz, C. D. Matzner, Mon. Not. R. Astron. Soc., {\bf 507}, 4335, (2021)
\bibitem[\protect\citeauthoryear {Schneider et al.}{2022}]{Schneider_ea_2022}
N. Schneider, V. Ossenkopf-Okada, S. Clarke, R. S. Klessen, S. Kabanovic, T. V. Veltchev, S. Bontemps, S. Dib, T. Csengeri, et al., Astron. \& Astrophys., {\bf 666}, 52 (2022)
\bibitem[\protect\citeauthoryear{Li}{2018}]{Li_18}
G.-X. Li, Mon. Not. R. Astron. Soc., {\bf 477}, 4951 (2018)
\bibitem[\protect\citeauthoryear{Donkov \& Stefanov}{2019}]{DS_19}	
S. Donkov, I. Zh. Stefanov, Mon. Not. R. Astron. Soc., {\bf 485}, 3224 (2019)
\bibitem[\protect\citeauthoryear{V\'azquez-Semadeni et al.}{2019}]{VS_ea_19}
E. V\'azquez-Semadeni, A. Palau, J. Ballesteros-Paredes, G. G\'omez, M. Zamora-Aviles, Mon. Not. R. Astron. Soc., {\bf 490}, 3061 (2019)
\bibitem[\protect\citeauthoryear{Donkov et al.}{2024}]{DSVK_24}	
S. Donkov, I. Zh. Stefanov, T. V. Veltchev, R. S. Klessen, Mon. Not. R. Astron. Soc., {\bf 527}, 2790 (2024)
\bibitem[\protect\citeauthoryear{Federrath \& Klessen}{2013}]{FK_13}
C. Federrath, R. S. Klessen, The Astrophys. Journal, {\bf 763}, 51 (2013)
\bibitem[\protect\citeauthoryear{Donkov, Veltchev \& Klessen}{2017}]{DVK_17}	
S. Donkov, T. V. Veltchev, R. S. Klessen, Mon. Not. R. Astron. Soc., {\bf 466}, 914 (2017)
\bibitem[\protect\citeauthoryear{Larson}{1969}]{Larson_69}
R. Larson, Mon. Not. R. Astron. Soc., {\bf 145}, 271 (1969)
\bibitem[\protect\citeauthoryear{Penston}{1969}]{Penston_69a}
M. V. Penston, Mon. Not. R. Astron. Soc., {\bf 145}, 457 (1969)
\bibitem[\protect\citeauthoryear{Shu}{1977}]{Shu_77}
F. H. Shu, The Astrophys. Journal, {\bf 214}, 488 (1977)
\bibitem[\protect\citeauthoryear{Hunter}{1977}]{Hunter_77}
C. Hunter, The Astrophys. Journal, {\bf 218}, 834 (1977)
\bibitem[\protect\citeauthoryear{Whitworth \& Summers}{1985}]{Whitworth_Summers_85}
A. Whitworth, D. Summers, Mon. Not. R. Astron. Soc., {\bf 214}, 1 (1985)
\bibitem[\protect\citeauthoryear{Mac Low \& Klessen}{2004}]{Mac Low_Klessen_2004}
M-M. Mac Low, R. S. Klessen, Reviews of Modern Physics, {\bf 76}, 125 (2004)
\bibitem[\protect\citeauthoryear{Naranjo-Romero, V\'azquez-Semadeni  \& Loughnane}{2015}]{N-R_ea_15}
R. Naranjo-Romero, E. V\'azquez-Semadeni, R. M. Loughnane, The Astrophys. Journal, {\bf 814}, 48 (2015)
\bibitem[\protect\citeauthoryear{Donkov \& Stefanov}{2018}]{DS_18}	
S. Donkov, I. Zh. Stefanov, Mon. Not. R. Astron. Soc., {\bf 474}, 5588 (2018)
\bibitem[\protect\citeauthoryear{Donkov et al.}{2022}]{DSVK_22}	
S. Donkov, I. Zh. Stefanov, T. V. Veltchev, R. S. Klessen, Mon. Not. R. Astron. Soc., {\bf 516}, 5726 (2022)
\bibitem[\protect\citeauthoryear {Jaupart \& Chabrier}{2020}]{Jaupart_Chabrier_2020}
E. Jaupart, G. Chabrier, The Astrophys. Journal L, {\bf 903}, L2 (2020)
\bibitem[\protect\citeauthoryear{Gomez et al.}{2021}]{Gomez_ea_2021}
G. G\'omez, E. V\'azquez-Semadeni, A. Palau, Mon. Not. R. Astron. Soc., {\bf 502}, 4963 (2021)
\bibitem[\protect\citeauthoryear{Mueller et al.}{2002}]{Mueller_ea_02}
K. E. Mueller, Y. L. Shirley, N. J. Evans II, H. R. Jacobson, The Astrophys. Journal S, {\bf 143}, 469 (2002)
\bibitem[\protect\citeauthoryear{Evans}{2003}]{Evans_03}
N. J. Evans II, in L. C. Curry, M. Fich, eds, SFChem: Chemistry as a Diagnostic of Star Formation. p. 157 (2003)
\bibitem[\protect\citeauthoryear{Wyrowski et al.}{2012}]{Wyr_ea_12}
F. Wyrowski, R. Gusten, K. M. Menten, H. Wiesemeyer, B. Klein, Astron. \& Astrophys., {\bf 542}, L15 (2012)
\bibitem[\protect\citeauthoryear{Wyrowski et al.}{2016}]{Wyr_ea_16}
F. Wyrowski, et al., Astron. \& Astrophys., {\bf 585}, A149 (2016)
\bibitem[\protect\citeauthoryear{Palau et al.}{2014}]{Palau_ea_14}
A. Palau, et al., The Astrophys. Journal, {\bf 785}, 42 (2014)
\bibitem[\protect\citeauthoryear{Csengeri et al.}{2017}]{Csengeri_ea_17}
T. Csengeri, et al., Astron. \& Astrophys., {\bf 600}, L10 (2017)
\bibitem[\protect\citeauthoryear{Zhang \& Li}{2017}]{Zhang-Li_17}
C.-P. Zhang, G.-X. Li, Mon. Not. R. Astron. Soc., {\bf 469}, 2286 (2017)
\bibitem[\protect\citeauthoryear{Bonnor}{1956}]{Bonnor_56}
W. B. Bonnor, Mon. Not. R. Astron. Soc., {\bf 116}, 351 (1956)
\bibitem[\protect\citeauthoryear{Ebert}{1957}]{Ebert_57}
R. Ebert, Z. Astrophys., {\bf 42}, 263 (1957)
\bibitem[\protect\citeauthoryear{Zhivkov}{1999}]{Zhivkov_99}
A. Zhivkov, Differential Equations Problems. Demokratichni tradicii - Demetra, Sofia (1999)
\bibitem[\protect\citeauthoryear{Riley, Hobson \& Bence}{2006}]{RHB_2006}
K. F. Riley, M. P. Hobson, S. J. Bence, Mathematical Methods for Physics and Engineering, Cambrige University Press, Cambridge, New York (2006)
\bibitem[\protect\citeauthoryear{Ballesteros-Paredes et al.}{2020}]{BP_ea_20}
J. Ballesteros-Paredes, et al., Space Science Reviews, {\bf 216}, (2020)
\bibitem[\protect\citeauthoryear{Wang et al.}{2024}]{Wang_ea_24}
C. Wang, et al., Astron. \& Astrophys., {\bf 681}, 33 (2024)

	
	
\end{thebibliography}
\end{document}